\documentclass{article}

\usepackage[preprint]{froggy_2025}
 
\usepackage{wrapfig}
\usepackage[utf8]{inputenc} 
\usepackage[T1]{fontenc}    
\usepackage{url}            
\usepackage{booktabs}       
\usepackage{amsfonts}       
\usepackage{nicefrac}       
\usepackage{microtype}      
\usepackage{xcolor}         

\usepackage{amsmath,amsfonts,bm}
\usepackage{xspace}

\def\eqref#1{equation~\ref{#1}}

\def\1{\bm{1}}

\DeclareMathAlphabet{\mathsfit}{\encodingdefault}{\sfdefault}{m}{sl}
\SetMathAlphabet{\mathsfit}{bold}{\encodingdefault}{\sfdefault}{bx}{n}

\newcommand{\done}{\textsc{R2E-Gym}\xspace}
\newcommand{\dtwo}{\textsc{SWE-Smith}\xspace}
\newcommand{\dthree}{\textsc{BugInstruct}\xspace}
\newcommand{\dfour}{\textsc{FeatAdd}\xspace}
\newcommand{\basemix}{\textsc{BaseMix}\xspace}
\newcommand{\bugpilot}{\textsc{BugPilot}\xspace}
\newcommand{\alldata}{\textsc{AllData}\xspace}

\newcommand{\frogboss}{\textsc{FrogBoss}\xspace}
\newcommand{\frogmini}{\textsc{FrogMini}\xspace}

\definecolor{darkgreen}{HTML}{005e19}
\definecolor{darkred}{HTML}{C00000}

\newcommand{\claude}{\textrm{Claude Sonnet 4}\xspace}
\newcommand{\opus}{\textrm{Claude Opus 4}\xspace}
\newcommand{\gptiv}{\textrm{GPT-4o}\xspace}
\newcommand{\gptv}{\textrm{GPT-5}\xspace}

\newcommand{\sbv}{SWE-Bench Verified\xspace}
\newcommand{\rtwoe}{R2E-Gym\xspace}
\usepackage{algorithm}
\usepackage[noend]{algpseudocode}
\usepackage{xcolor}
\usepackage{booktabs}
\usepackage{threeparttable}
\definecolor{darkGreen}{rgb}{0.2,0.5,0.2}
\definecolor{mydarkblue}{rgb}{0,0.08,0.45}
\usepackage[colorlinks,citecolor=mydarkblue,urlcolor=mydarkblue,linkcolor=mydarkblue]{hyperref}
\usepackage[capitalize,noabbrev]{cleveref}
\usepackage{url}
\usepackage{graphicx}
\usepackage{subcaption}
\usepackage{enumitem}
\usepackage{booktabs,threeparttable}
\usepackage{tikz}           
\usepackage{multirow}
\usepackage{mathtools}

\usetikzlibrary{patterns,patterns.meta,positioning,backgrounds,calc}

\usepackage{todonotes}
\usepackage[normalem]{ulem}

\makeatletter
\newcommand*\iftodonotes{\if@todonotes@disabled\expandafter\@secondoftwo\else\expandafter\@firstoftwo\fi}  
\makeatother

\title{BugPilot: Complex Bug Generation for \\Efficient Learning of SWE Skills}

\author{
\Authfont{Atharv Sonwane$^{*1}$, Isadora White$^{*2}$, Hyunji Lee$^{3}$,}\\
\vspace{-0.2em} 
\Authfont{Matheus Pereira$^{4}$, Lucas Caccia$^{4}$, Minseon Kim$^{4}$, Zhengyan Shi$^{4}$, Chinmay Singh$^{4}$,}\\
\Authfont{Alessandro Sordoni$^{4}$, Marc-Alexandre Côté$^{4}$, Xingdi Yuan$^{4}$}\\
\vspace{0.2em}
\Affilfont{$^{*}$Equal contribution\:\:\:\:$^{1}$Cornell University\:\:\:\:$^{2}$University of California San Diego} \\
\Affilfont{$^{3}$University of North Carolina at Chapel Hill\:\:\:\:$^{4}$Microsoft Research} \\
\vspace{0.2em}
\Affilfont{ays57@cornell.edu\:\:\:\:i2white@ucsd.edu\:\:\:\:debug-gym@microsoft.com}\\
\vspace{0.2em}
\textcolor{froggy-green}{\Authfont{https://microsoft.github.io/debug-gym/}}\\
}

\begin{document}

\maketitle

\begin{abstract}
High quality bugs are key to training the next generation of language model based software engineering (SWE) agents.
We introduce a novel method for synthetic generation of difficult and diverse bugs.
Our method instructs SWE Agents to introduce a feature into the codebase whereby they may unintentionally break tests, resulting in bugs. 
Prior approaches often induce an out-of-distribution effect by generating bugs \emph{intentionally} (e.g. by introducing local perturbation to existing code), which does not reflect realistic development processes.
We perform qualitative analysis to demonstrate that our approach for generating bugs more closely reflects the patterns found in human-authored edits.
Through extensive experiments, we demonstrate that our bugs provide more efficient training data for supervised fine-tuning, outperforming other bug datasets by 2\% with half the training data (1.2k vs. 3k bugs).
We train on our newly generated bugs in addition to existing bug datasets to get \frogboss{} a state-of-the-art 32B parameter model on \sbv with a pass@1 of 54.6\% and \frogmini{} a state-of-the-art 14B model on \sbv with a pass@1 of 45.3\% on \sbv averaged over three seeds.
\end{abstract}

\begin{figure}[h]
    \centering
    \includegraphics[width=0.90\linewidth]{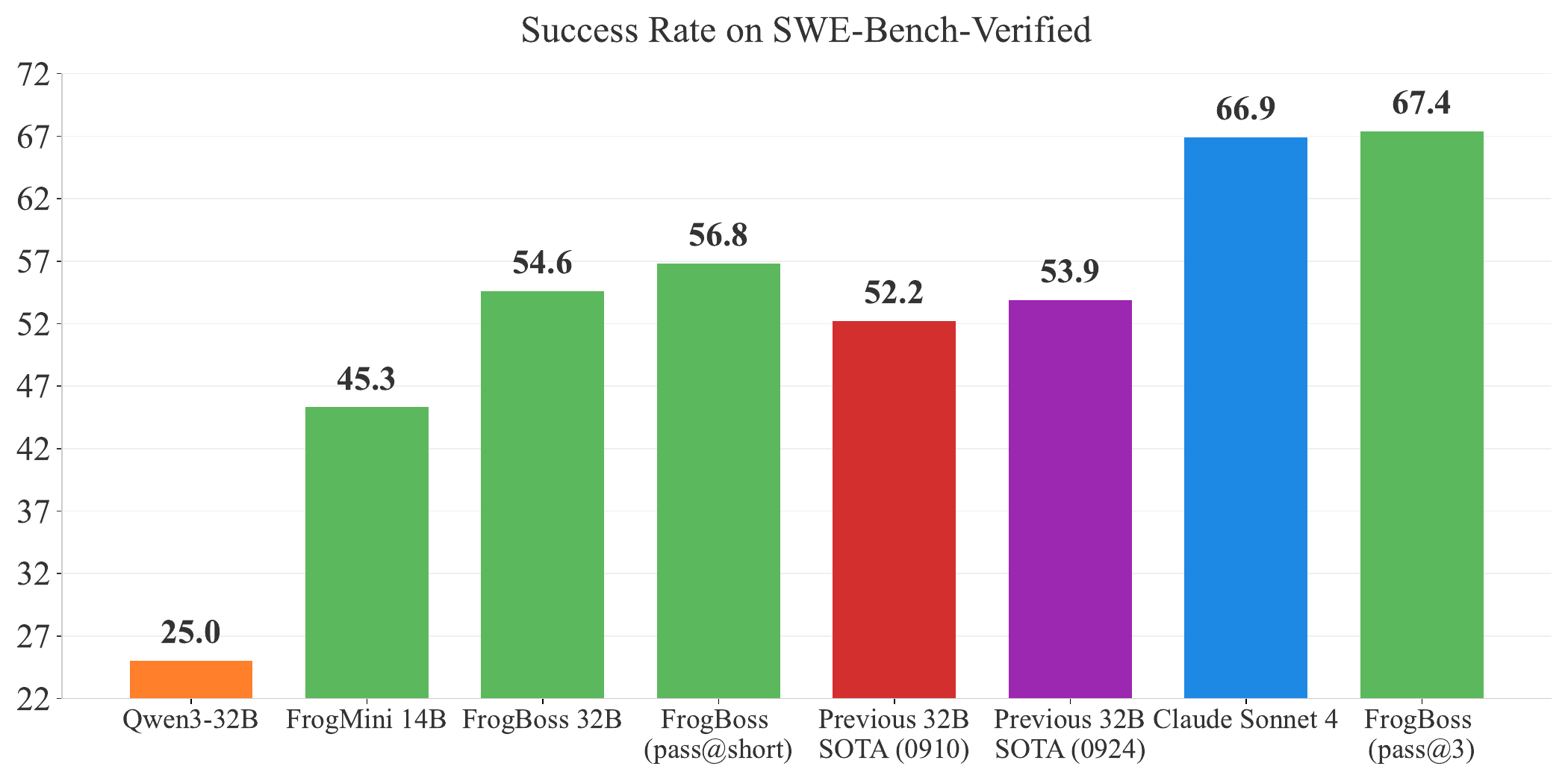}
    \caption{\textbf{Comparison to previous SoTA results.} We train \frogboss{} with our collected data including our new \dfour datasets, and \frogboss{} achieves 54.6\% pass@1 averaged over three seeds. With pass@3 we achieve a score of 67.4\%, outperforming \claude, illustrating the performance gains we could get with a good verifier. The minimal test time scaling refers to the strategy of selecting the shortest trajectory among three rollouts. Pass@short refers to taking the shortest rollout from a group of three rollouts on a given problem.}
    \label{fig:sota_figure}
\end{figure}

\section{Introduction}

Large language model (LLM)–based agents have recently made strong progress on software engineering (SWE) tasks~\citep{NEURIPS2024_5a7c9475, jimenez2023swe-bench, pan2024training, wei2025swe}. 
However, the strongest agents rely on proprietary models, and improving open-weight models on these tasks remains challenging. 
Training models to solve bugs using either supervised fine-tuning from expert data or reinforcement learning is promising~\citep{yang2025swe, jain2025r2e, deepswe2025, wei2025swe}. 
However scaling this approach requires large, high-quality bug datasets.

Existing bug curation strategies fall into two camps. 
One mines real bugs from pull requests and commits in open-source repositories, which demands careful issue localisation and filtering~\citep{xie2025swe, badertdinov2025swe, pan2024training, wang2025swe, zhang2025swe}. 
Alternatively, synthetic bug generation injects faults into existing codebases, allowing researchers to scale data without being bottlenecked by the availability of existing commits, pull requests or issues~\citep{yang2025swe}. 
A notable example of synthetic bugs is SWE-Smith~\citep{yang2025swe}, which relies on hand-engineered rules and LLM re-implementations of existing functions in order to perturb the codebase until tests break. 
Although useful, this method produces datasets skewed toward a narrow set of bug types with fixes that are short and typically confined to a single file. 
This might undermine the transferability of models trained on such synthetic data to real-world scenarios, where bugs typically arise through natural development processes rather than deliberate injection of errors.



In this work, we introduce \textsc{BugPilot}, a novel approach to synthetic bug generation that leverages software engineering agents to create more naturalistic bugs through realistic development workflows. 
A naive approach to agentic bug generation would be what we refer to as \dthree: to explicitly instruct a SWE agent to \emph{intentionally} introduce a bug in an existing code-base.
This approach generates bugs that qualitatively do not resemble realistic bugs. 
Rather than intentionally injecting errors, our method tasks SWE agents with developing new features within existing repositories (which we refer to as \dfour). 
This results in naturally introduced bugs when these implementations \emph{unintentionally} break existing test suites.
We detect when such breakages arise and record the state of the repository at this point as containing a bug that needs to be resolved.
This process mirrors authentic software development scenarios where bugs commonly arise as unintended side effects of feature development and code modifications.

\begin{figure}[t]
    \centering
    \includegraphics[width=0.9\linewidth]{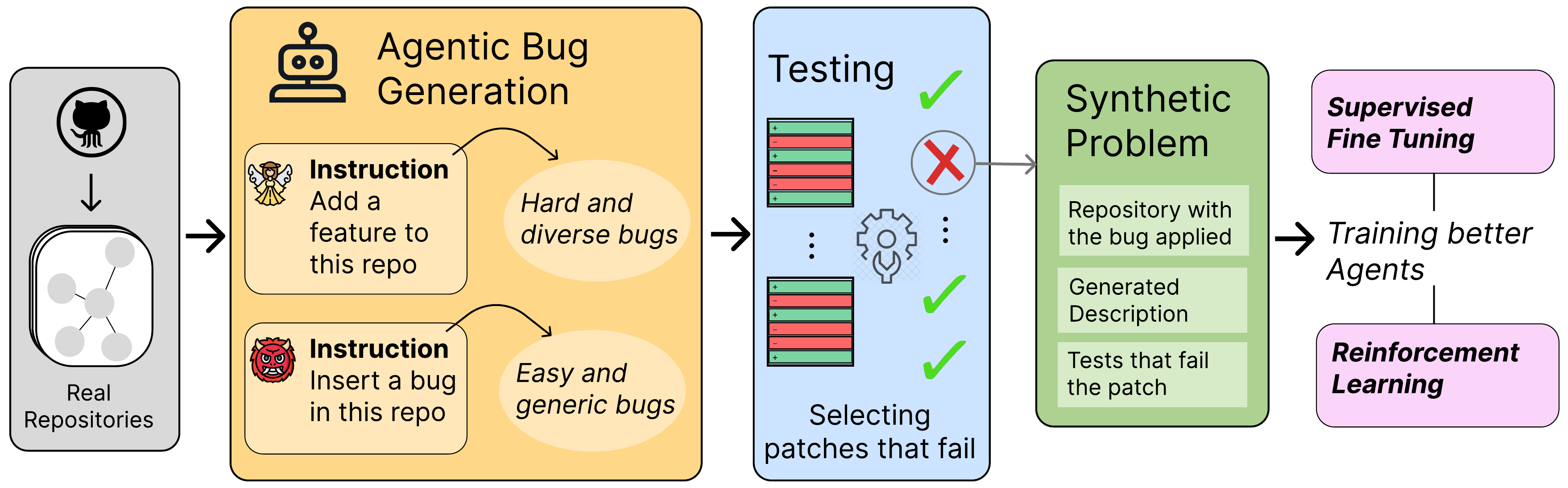}
    \caption{\textbf{Illustration of our \bugpilot pipeline.} First, we instruct SWE Agent \citep{NEURIPS2024_5a7c9475} with \claude to introduce bugs, either through deliberate attempts or by attempting to add a feature. Then, we check whether these modifications resulted in the tests for the repository failing. If the tests fail, then we add this to our dataset of bugs. Otherwise, we ask the model to continue changing the code until the tests fail.}
    \label{fig:bug_pilot}
\end{figure}

Through qualitative and quantitative analyses, we demonstrate that our generated bugs are not only more challenging for current agents, but are more diverse and exhibit more natural characteristics compared to existing synthetic datasets.
Comparing \textit{unintentionally} generated bugs (\dfour) to \textit{intentionally} generated bugs (\dthree and \dtwo) for fine-tuning of a base model reveals that unintentional bugs provide much more efficient training examples - performing 2\% better with half the number of training trajectories (1.2k vs. 2.3k).
Using our bugs to train an agent with reinforcement learning, our model achieves 52.4\% on \sbv with a 32B parameter model (Pass@1 averaged across 3 seeds).
When training using an extended set of all our collected data, we achieve state-of-the-art results for a 32B model with 54.6\% on SWE-Bench Verified.

Our contributions are fourfold: (1) we propose \bugpilot, a novel methodology for generating synthetic bugs through realistic development workflows with SWE agents (\Cref{fig:bug_pilot}), (2) through qualitative analysis we categorise bug datasets and show that bugs generated through \dfour reflect a more natural category distribution (\Cref{sec:bug_analysis}), (3) we demonstrate that \textit{unintentionally} generated bugs provide more efficient training data for supervised fine-tuning and reinforcement learning than \textit{intentionally} generated bugs, (4) we train \frogboss{}\footnote{Who loves to eat bugs! \includegraphics[height=10pt,trim=0 2cm 0 0cm]{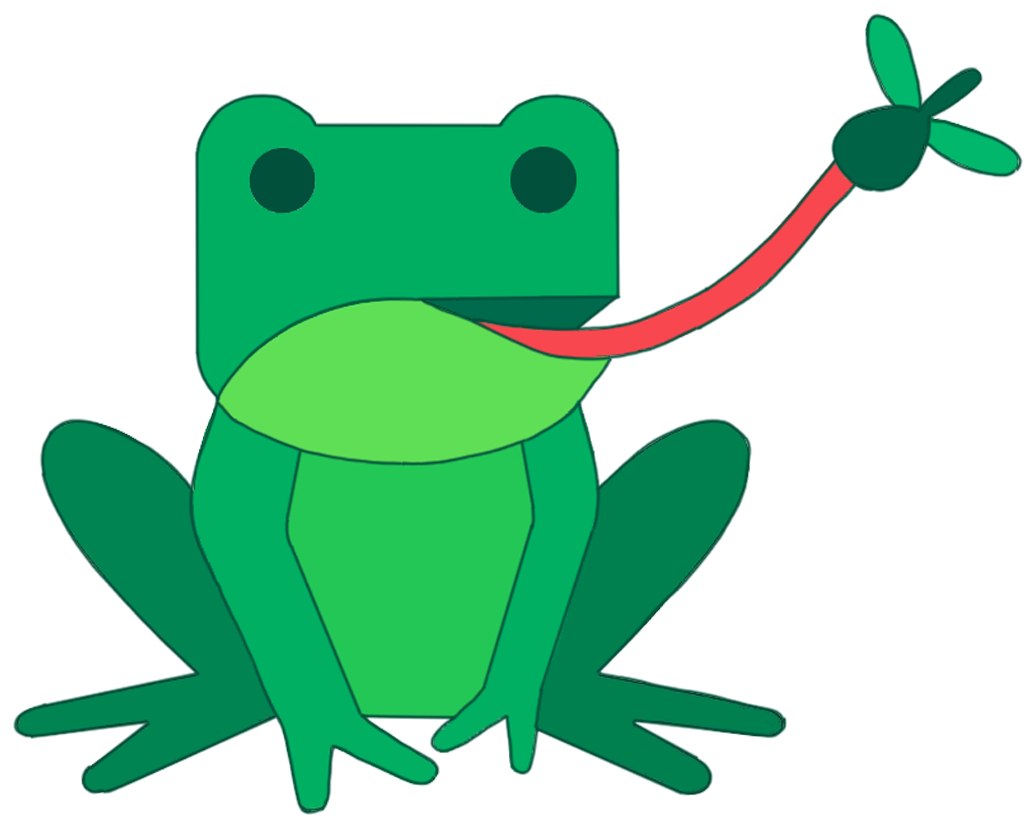}}, a 32B model with 54.9\% Pass@1 averaged over three seeds and \frogmini{}, a 14B model with 45.3\% averaged over three seeds by training with on a dataset combining all of our unintentionally generated \dfour bugs with previous bug datasets for three epochs.



\section{Related Work}


\paragraph{Software Engineering Benchmarks and Tasks} SWE-Bench \citep{jimenez2023swe-bench} introduced $2,294$ problems from real GitHub issues and the corresponding solution PR, driving the first benchmark to study whether state of the art LLM agents can solve real-world SWE tasks. 
However, this initial set of problems has a few issues: (1) the solvability of these tasks, (2) the relative complexity of these tasks, and (3) the limited number of bugs. 
To solve the former, \sbv was introduced: a set of tasks verified by human engineers of whether they were solvable given the information described in the problem statement. 
To address the issue of a limited number of bugs, SWE-Fixer introduced a dataset of 110k human-authored bugs extracted from GitHub Issues~\citep{xie2025swe}, SWE-rebench~\citep{badertdinov2025swe} introduced a set of 60k human generated tasks, SWE Gym introduced 2.4k new tasks~\citep{pan2024training}, SWE-bench-Live~\citep{zhang2025swe} introduced 1.5k tasks covering 164 repositories. Recent work has studied the synthetic generation of bugs in the form of SWE-Smith \citep{yang2025swe} where LLMs rewrite functions to add bugs and \rtwoe \citep{jain2025r2e} where bugs are extracted from the commits of repositories rather than GitHub issues. 

\paragraph{Frameworks for Software Engineering Agents} Most work revolving around the development of improved software development frameworks involves improving the agentic framework and tools surrounding the base LLM agent. 
There have been many agentic frameworks that subsequently improve on SWE Bench performance such as \citep{NEURIPS2024_5a7c9475, ma2024lingma, yuan2025debug, wang2024openhands, jain2025r2e} implementing new tools such as a pdb debugger in Debug Gym, or complex navigation and manipulation tools in Moatless tools.
In contrast to agentic approaches are pipeline based approaches such as Agentless~\citep{xia2024agentless}, a framework where instead of relying on an end-to-end agent loop, the tool goes through three pre-set phases: localization, repair, and patch validation. 

\paragraph{Learning Approaches for Developing Better SWE Agents} A number of training frameworks have been introduced to train better SWE agents using supervised fine-tuning and reinforcement learning \citep{Luo_2025, wei2025swe}.
Current learning paradigms that have been attempted include SFT, performed with SWE-Smith bugs, \rtwoe bugs, and SWE-gym bugs as well as RL performed with \rtwoe bugs and RL with human-authored bugs \citep{wei2025swe}.
Moreover, approaches such as test-time scaling using test generation or LLM-as-a-judge yield significant performance improvements in both SWE-Gym and \rtwoe, including using Monte Carlo tree search (MCTS) \citep{antoniades2024swe} or inference time process reward models \citep{gandhi2025agents} to guide better search procedures with verification. 


\section{Automatic Bug Generation} \label{sec: bug_generation}

\subsection{Background}

A \textit{bug} consists of (1) a repository with some code that functions incorrectly, (2) a natural language description of how this bug manifests while using the code and (3) tests that can be used to verify if the bug has been fixed.
In the context of \textit{bug-fixing} tasks, agents are provided with (1) the buggy repository and (2) the problem description and are expected to produce code changes in the repository.
These code changes are said to be correctly resolving the bugs if they result in the (3) verification test cases passing.

Collections of bugs, such as SWE-Bench, serve dual purposes in advancing software engineering agents. 
First, they enable systematic evaluation on realistic debugging scenarios. 
Second, they provide valuable training data: bugs can be used to generate expert trajectories for supervised fine-tuning, or serve as RL environments for bug-fixing skills.

Bug datasets are traditionally curated from open-source repository histories. 
However, this approach requires substantial effort to filter suitable examples and is fundamentally constrained by the availability of development histories on platforms like GitHub.
Alternatively, synthetic bug generation pipelines enable bug creation for arbitrary repositories and programming languages without depending on existing commit histories.
Before introducing our agentic bug generation approach, we briefly describe \done~\citep{jain2025r2e} and \dtwo~\citep{yang2025swe}:


\paragraph{Human-authored Edits.}
The \done dataset constructs bugs by reverting commits from Python repositories. 
Commits are chosen using heuristics such as edit size. 
Verification tests are extracted by identifying tests that fail on the buggy version but pass on the fixed version. 
Problem statements for each bug are generated by prompting an LLM with the code diff and test failures.

\paragraph{Synthetic Bug Generation.}
The \dtwo dataset introduces synthetic bugs into Python repositories through three mechanisms: (1) procedural code modifications, (2) LLM-based reimplementation of individual functions, and (3) PR inversion, where an LLM attempts to reverse existing pull requests by regenerating files that undo the changes. All approaches verify bugs using failing test cases, with problem descriptions generated via LLM prompting.

\subsection{BugPilot: Agentic Generation of Bugs} \label{sec: bug-pilot}

\begin{figure}[t]
    \centering
    \includegraphics[width=0.8\linewidth]{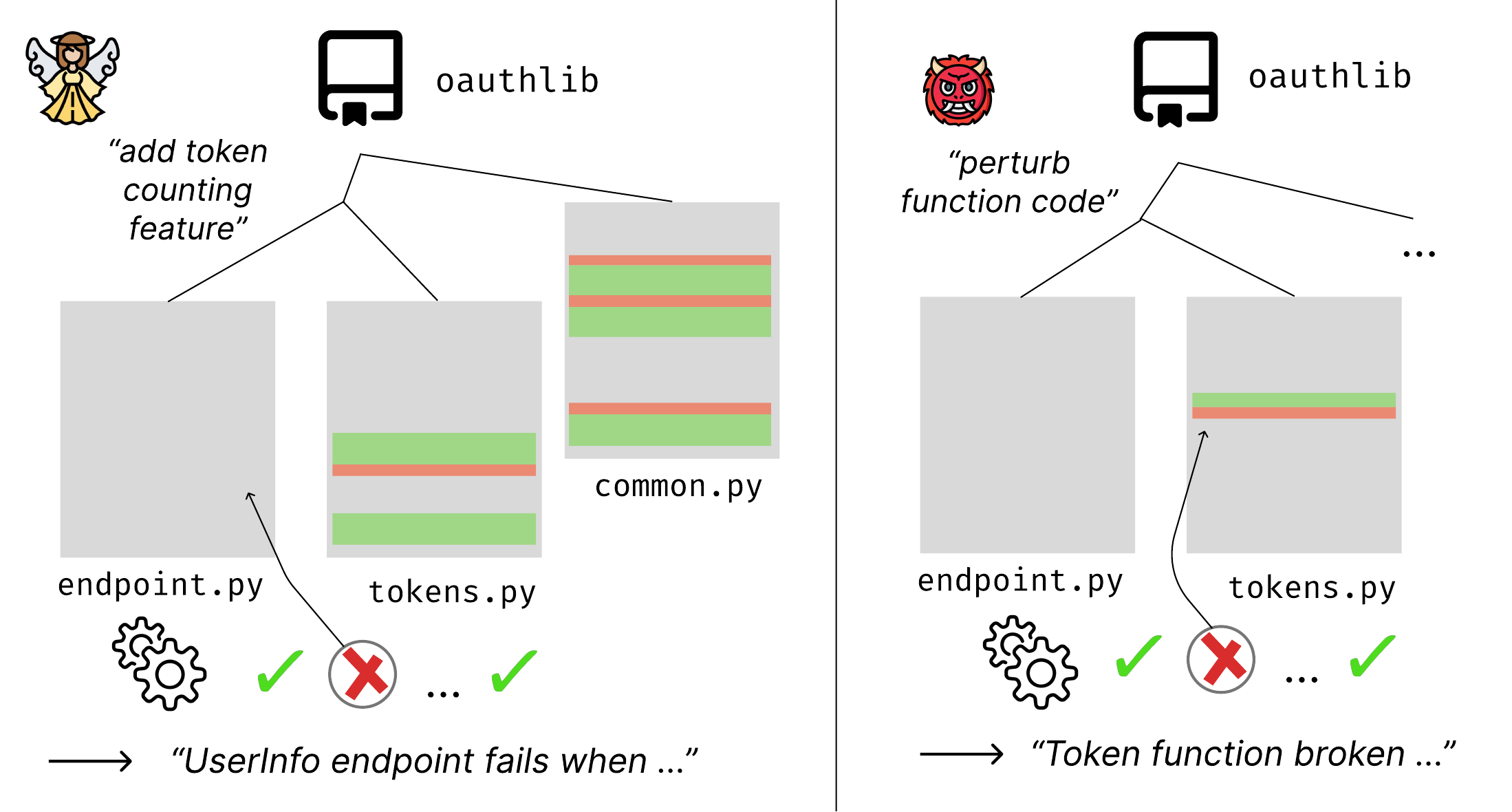}
    \caption{\textbf{Contrasting approaches to adding bugs}. On the left, our \dfour approach first attempts to implement a token counting feature in the repository. This results in large changes across multiple files as well as a test case failure arising from a seemingly unrelated part of the repository. In contrast on the right, approaches like \dtwo and our proposed baseline \dthree make local perturbations to the code which cause related tests to fail. We can see that \dfour more closely resembles how bugs arise during the development process, where test failures can occur due to complex interactions between changes.}
    \label{fig:bug_comparison}
\end{figure}

Current methods for generating synthetic bugs (e.g. SWE-Smith) work by perturbing the code until the tests break.
We hypothesize that SWE agents themselves might be used to introduce bugs in a way more reflective of real-life software engineering.
We start with the set of 128 SWE-Smith repositories where, for each repository we have a containerised environment (docker image) where the codebase along with all dependencies have been installed.
To synthesize bugs within these repositories we use SWE-Agent~\citep{NEURIPS2024_5a7c9475} with \claude. 
The \textit{agent} here is a system that interacts with these containers and ultimately makes a set of changes to files across the repository.
It consists of a loop that prompts the language model with high level instructions along with outputs from the last step.
The language model is asked to generate a tool call, where possible tools in our configuration include viewing / editing files as well as executing terminal commands in the container.
The loop terminates after the language model generates the {\tt submit} tool or if certain limits are reached.
An illustration of our bug generation pipeline can be found in \Cref{fig:bug_pilot} and a comparison between the bugs generated using \dfour approach can be found in \Cref{fig:bug_comparison}.
We consider two ways to introduce bugs within this framework -

\paragraph{Intentional Bug Introduction (\dthree)} We instruct the agent to introduce bugs into the repository by enriching its system prompt with guidance on bug integration techniques and verification steps to confirm that changes break existing functionality.
However, this intentional approach produces bugs that lack diversity and are typically far simpler than real-world bugs, as we demonstrate later. 
This distributional mismatch ultimately limits their effectiveness for improving agentic coding performance. 
We refer to this method as \dthree.

\paragraph{Buggy Feature Addition (\dfour)} Many real world bugs in the software development process arise when existing code is modified incorrectly to support new features. 
We emulate this by tasking our agent to come up with and implement a new feature for the given repository while preserving existing functionality (detailed prompt provided in Appendix~\ref{sec:buggen_prompts}).
Whenever the feature breaks an existing test, a bug is created. Unlike the earlier approach, bugs here are \textit{unintentional} and thus are more likely to align to naturally occuring bugs. We refer to this method as \dfour.

For both of these approaches, given a repository, we execute both strategies multiple times in order to collect differing approaches from at performing the same task.
We evaluate whether a run resulted in a bug by running tests after the agent has submitted and making sure at least one test fails (see Figure~\ref{fig:bugcat}).
The description of the bug for each ``buggy'' run is generated by prompting the language model to generate a bug-report given the output of the failed tests, following SWE-Smith~\citep{yang2025swe}.
The final synthetically generated bug is thus composed of the code changes made by the agent during its execution along with the failing test and its outputs.
Our approach enjoys the same scalability benefits of SWE-smith, wherein once a repository is setup in a containerised environment, no additional manual effort is required in order to generate more bugs.

\section{Datasets and Training Methodology}
\label{sec:dataset}

Below we describe how we collect trajectories for training from various bug datasets, along with our methodology for training on these trajectories.

\begin{table*}[t]
\centering
\caption{\label{tab:difficulty_stats}\textbf{Solve statistics of \claude, \gptiv and \gptv using \rtwoe as the agentic scaffold.} Agentic bugs generated via either \dfour or \dthree are more difficult than the \dtwo and \done bugs, with \dfour bugs being the most difficult of them all across the three models.}
\begin{tabular}{lcccccc}
\toprule
\textbf{Models} & \done & \dtwo & \dthree & \dfour \\
\midrule
\underline{\claude} & 63.5\% & 65.9\%  & 54.6\% & 41.4\% \\
\;Successful Trajectories & 3,208 &  2,611 & 2,330 & 1,243 \\
\;Avg Steps & 42.0 & 39.2 & 43.1 & 45.5 \\
\;Avg Observation Tokens & 460.1 & 448.8 & 464.7 & 433.5 \\
\;Avg Assistant Content / Trajectory & 34.5 & 30.5 & 32.9 & 35.3 \\
\;Avg Assistant Content Tokens & 56.8 & 59.1 & 60.4 & 61.0 \\
\midrule
\underline{\gptiv} & 32.8\% & 29.4\% & 13.4\% & 18.5\% \\
\;Avg Assistant Content / Trajectory & 5.6 & 5.8 & 6.2 & 7.7 \\
\;Avg Assistant Content Tokens & 150.7 & 152.1 & 157.7 & 154.4 \\
\midrule
\underline{\gptv} & 68.7\% & 77.5\% & 67.8\% & 53.4\% \\
\;Avg Assistant Content / Trajectory & 0.8 & 0.5 & 12.3 & 14.5 \\
\;Avg Assistant Content Tokens & 465.6 & 481.3.6 & 21.0 & 22.0 \\
\bottomrule
\end{tabular}
\end{table*}

\paragraph{Agentic Framework}
For all our inference and training purposes we use R2EGym as our agentic scaffold, because of its previous usage and strong performance on \sbv \citep{jain2025r2e}. 
Moreover, the R2EGym scaffold is built into RLLM \citep{rllm2025}, a framework for training language models with RL on SWE tasks, making it convenient to compare SFT to RL. 
The R2EGym scaffold offers to the agent four tools: the \verb_file-editor_, \verb_execute-bash_, \verb_search_ and \verb_finish_. 

\paragraph{Supervised Fine-tuning} 
We collect trajectories for training with supervised fine-tuning (SFT) using \claude and bug datasets from all four of the bug generation techniques described above. 
To create the dataset for SFT, we use rejection sampling on each of the datasets.
We generate the trajectories using a 64k context length and 10k max prompt length and then filter them based on success. 
The statistics of these trajectories are reported in Table~\ref{tab:difficulty_stats}. 
For our student model, we choose Qwen3-32B~\citep{qwen3technicalreport}. 
We fully fine-tune our model using LlamaFactory~\citep{zheng2024llamafactory} with a learning rate of 1e-5, no weight decay, we perform 2 epochs of training with a maximum context length of 32k tokens. 
If a trajectory generated at 64k tokens is successful, we train on the first 32k tokens. 
This occupies one node of 8 Nvidia H100 for 10 hours.

\paragraph{Reinforcement Learning}
Recent work has shown promise in using Reinforcement Learning to fine-tune LLMs for tasks with verifiable rewards, especially for Maths and Competition programming.
Software development tasks as discussed in this paper differ from the above in that they are multi-turn, requiring the model to interact with the environment in a diverse way.
Following DeepSWE \citep{deepswe2025}, we employ the RLLM \citep{rllm2025} framework for fine-tuning language models using RL with various rewards. 
Similar to the SFT paradigm, we generate rollouts with a max context length of 64k tokens, but truncate to the first 32k tokens for training. 
To train reinforcement learning for 25 steps with 64 bugs per step and 8 rollouts per bug, we require 8 nodes of 8 $\times$ H100s for 50 hours. 

\paragraph{Evaluation} Evaluation was performed on \sbv and every model was run over three seeds with a context length of 64k, 100 max steps and a temperature of 1. Full hyper-parameters can be found in Appendix~\ref{sec:reinforcement_learning}.

\paragraph{Methodology and Data Mixtures}
Our main experiments use a base mixture of \done and \dtwo bugs with a total of 5,621 successful resolution trajectories from \claude (as reported in Table~\ref{tab:difficulty_stats}, we have 5,819 trajectories for these two datasets but we leave out 198 trajectories for validation). We call this mixture \basemix. 
We first fine-tune our base model on this mixture followed by perform another round of fine-tuning on 1.2k trajectories generated from our agentic generated \dthree and \dfour bugs. 
For a fair comparison, we perform this second stage of fine-tuning on additional 1.2k trajectories (same size as above datasets) from \done and \dtwo, not included in \basemix.
Finally, we experiment with fine-tuning  the base model on \alldata, consisting of \basemix, \dfour along with the 1k trajectories each from\done and \dtwo.




\section{Results}

\subsection{Bug Analysis} \label{sec:bug_analysis}


\begin{figure}[t]
\centering
\includegraphics[width=0.9\linewidth]{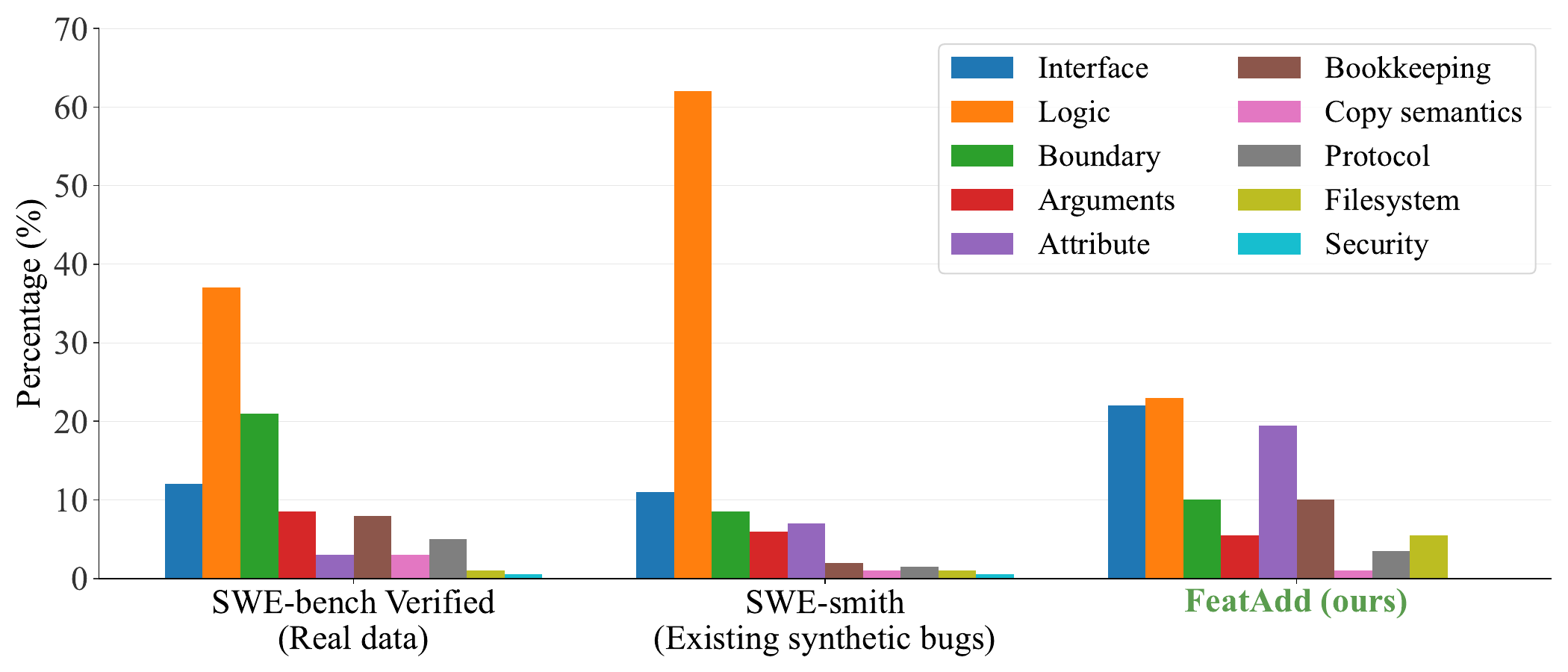}
\caption{\label{fig:bugcat}\textbf{Distribution of common bug types across different bug datasets.}
\dfour demonstrates the most even distribution of bugs compared to prior work as well as our agentic baseline \dthree.
\dtwo bugs shows a particular skew towards \textbf{logic} and conditional bugs.
This can be explained by the rule based and local nature of the \dtwo generation process.
Most bugs generated by \dthree are state consistency / \textbf{bookkeeping} / caching bugs or \textbf{copy semantics} / mutability aliasing / in-place mutation of inputs bugs. The distribution of \dfour bugs is similar to \done and \textsc{SWE-Bench}, which is closest to the human authored edit distribution found in real repositories.}
\end{figure}

\begin{table*}[t]
\centering
\caption{\textbf{Bug Statistics.} We compare bugs from different generation/collections methods. \sbv contains real-world bugs, \done uses human-authored edits, while others use synthetic generated bugs. \dfour characteristics differ significantly from previous approaches in that the patch used to introduce the bug has many more tokens and on average twice as many files changed.}
\begin{tabular}{@{}lccccccc@{}}
\toprule
Feature & SWE-B-V &\done & \dtwo & \dthree & \dfour \\ 
\midrule
Total tasks & 500 & 1000 &  1000 & 1000 & 785 \\
Problem tokens & 447.7 & 264.6 &  312.0  & 332.6 & 304.4 \\
Avg diff patch tokens & 394.0 & 352.6 & 598.2  & 435.3 & 4376.0 \\
Avg files modified & 1.2 & 2.6 & 1.2 & 1.3 & 4.2 \\
Avg net lines changed & 5.6 & 36.0 & -3.2 & 12.4 & 415.9 \\
Unique repositories & 12 & 10  & 125  & 111 & 86 \\
Avg tasks per repo & 41.7 & 100.0  & 8.0& 9.0 & 9.1 \\
Claude resolve rate & 70.8 & 63.5 & 65.9 & 54.6 & 41.4 \\
\bottomrule
\end{tabular}
\label{tab:bug_analysis}
\end{table*}

We first study the characteristics of the bugs generated by our approach by categorizing them and comparing the distribution of bug-types with both human-authored bugs (\sbv and \done) and AI-generated (\dtwo).

\textbf{Bug Categorization} In Figure~\ref{fig:bugcat} we present results from an LLM-aided categorisation of bugs into common bug-types.
Bugs generated using \dfour demonstrate a more even distribution of bugs across various categories compared to prior work, which is skewed to a few bug types.
This demonstrates that \dfour bug generation approach can be used to generate diverse bugs synthetically, which match real world bug distributions, unlike prior synthetic generation baselines.
Further details of the bug categorisation can be found in  Appendix~\ref{sec:appendix_bugcat} and examples of \dfour bugs of different types can be found in Appendix~\ref{sec:featadd_examples}. 

\textbf{Bug Statistics} In Table~\ref{tab:bug_analysis}, we study how the patch that introduces the bugs differs quantitatively across different generation methods. 
\dfour results in starkly different bug patches to the other approaches, with the changes usually being made across multiple files and of a greater magnitude. 

\textbf{Bug Difficulty} We evaluate bug difficulty by measuring the ability of a strong coding agent to solve generated bugs from that dataset. 
We use the \rtwoe \emph{scaffold}~\citep{jain2025r2e} as our agentic scaffold and we use \claude as our strong coding LLM. 
For all bugs, we sample 4 attempts. 
The results in Table~\ref{tab:difficulty_stats} show that \dfour bugs are the most challenging even for frontier models and the solve success rate drops from 63.5\% to 41.4\%.

\subsection{Training on Bugs}

\begin{table}[t]
    \small
    \centering
    \caption{\textbf{Comparison to Current State-of-the-Art.} 
    We achieve state-of-the-art results for a 14B and 32B model by supervised fine-tuning on bugs generated by \dfour in addition to existing bug datasets (\alldata).
    We also show that training with reinforcement learning (RL) on \dfour for only 25 steps; starting from a base model fine-tuned with \basemix can outperform previous state of the art with less that half the amount of trajectories.
    Our model trained with supervised fine-tuning (SFT) on \basemix+ \dfour achieves near state-of-the-art results with 40\% of the total training trajectories (7k vs. 12k) and 5\% of the total bugs (3k vs. 60k).}
    \begin{tabular}{llccc}
        \toprule
         \textbf{Model/Method }& \textbf{Scaffold} & \textbf{Bugs} & \textbf{Trajectories} & \textbf{SWE-Bench (V)}  \\
        \midrule
        \textit{Proprietary Models} & & & \\
        Claude Sonnet 4        & Moatless Tools & - & - & 70.8 \\
        Claude Sonnet 4 & SWE-Agent      & - & - &  66.6 \\
        Claude Sonnet 4 & \rtwoe        & - & - & 66.9\\
        GPT-4o          & \rtwoe        & - & - & 29.3 \\
        GPT-5          & \rtwoe        & - & - & 65.7 \\
        \midrule 
        \textit{Open Weights Models} \\
        DeepSeek-R1-0528 \citep{guo2025deepseek} & OpenHands & - & - & 45.6\\
        Qwen3-Coder-480B \citep{qwen3technicalreport} & mini-SWE-Agent & - & - & 55.4 \\
        GLM-4.5-358B \citep{zeng2025glm}     & SWE-Agent      & - & - & 64.2\\
        GLM-4.5-358B \citep{zeng2025glm}  & mini-SWE-Agent & - & - & 54.2 \\
        SWE-Fixer-72B \citep{xie2025swe} & SWE-Fixer & 110k & - & 32.8\\
        SWE-RL-70B \citep{wei2025swe} & Agentless & & & 41.0 \\
        
        CWM-32B \citep{cwm2025}  & Agentless & - & - & 53.9\\
        DeepSWE-32B-Preview \citep{deepswe2025} & \rtwoe & 4.6k & - & 42.2 \\
        
        SWE-Gym-32B \citep{pan2024training} & OpenHands & 2.4k & 491 & 20.6 \\
        
        R2E-Gym-32B \citep{jain2025r2e} & \rtwoe & 4.6k & 4.5k & 34.4\\
        
        Skywork-SWE-32B \citep{zeng2025skywork} & OpenHands & 10.1k & 8k & 38.0 \\
        
        SWE-Smith-LM-32B \citep{yang2025swe} & SWE-Agent & 50k & 5k & 40.2\\
        SWE-Mirror-LM-32B \citep{wang2025swe} & OpenHands & 60k & 12k & 52.2 \\
        
        \midrule 
        \textit{Ours (14B)} \\
        \basemix+ \dfour 14B (SFT) & \rtwoe & 3k & 7k & \textbf{41.1} \\
        \basemix+ \dfour 14B (RL) & \rtwoe  & 3k & 5.8k (\basemix) & \textbf{42.2} \\
        \frogmini{} (All Data, SFT)        & \rtwoe  & 3k & 9k & \textbf{45.3} \\
        \midrule 
        \textit{Ours (32B)}\\
        \basemix+ \dfour 32B (SFT) & \rtwoe  & 3k & 7k & \textbf{51.9}\\
        \basemix+ \dfour 32B (RL) & \rtwoe & 3k & 5.8k (\basemix) & \textbf{52.4}\\
        \frogboss{} (All Data, SFT) & \rtwoe & 3k & 9k & \textbf{54.6} \\
        
        \bottomrule
        
    \end{tabular}
    \label{tab: sota-comparison}
\end{table}

Table~\ref{tab: sota-comparison} presents the performance of our Qwen3-32B models against prior work, comparing three training configurations: RL fine-tuning on \dfour bugs, supervised fine-tuning on mixed trajectory datasets, and our best approach training on a combined dataset.
Our RL-trained model establishes a new state-of-the-art on \sbv, achieving 52.4\% Pass@1 (averaged over three seeds). 
Notably, in the supervised fine-tuning setting, \basemix+ \dfour reaches nearly identical performance at 51.9\% while being substantially more data-efficient than concurrent work—using 40\% fewer trajectories and only 5\% as many bugs in the SFT dataset compared to SWE-Mirror~\citep{wang2025swe}.
Building on these results, we trained our best-performing model, \frogboss{}, using the full trajectory set: 2k bugs each from \done, \dtwo, and \dfour. Despite using a 25\% smaller dataset than previous work (9k vs. 12k bugs), this configuration achieves 54.6\% Pass@1, our highest result.
These findings demonstrate that \dfour, our high-quality synthetic dataset, enables both superior performance and greater data efficiency in training software engineering agents.

We also present results from the same set of experiments as above using Qwen3-14B base model in Table~\ref{tab: sota-comparison}.
We name the best performing 14B model as \frogmini{}. 
Our experiments with the smaller models demonstrate the same trend as in the 32B model setting: a model fine-tuned on the \basemix trajectories can be further improved by both SFT and RL training using the \dfour data.
We also observe that SFT on the the full set of trajectories (as the ones used to train \frogboss{}) results in best performance with \frogmini{}.
These results on the smaller model further support the idea that high quality data from using \dfour can lead to improved bug-fixing performance more generally.


\paragraph{Comparison to continued fine-tuning on other synthetic data mixtures.} 
In Table~\ref{tab:main_result_32b} we compare \dfour with \done, \dtwo and \dthree in the context of data mixtures used for fine-tuning.
We compare models initially trained on \basemix (5.6k trajectories from previously available bugs) followed by additional data.
We find that \basemix + \dfour (1.2k trajectories from a strong teacher model solving our synthetically generated bugs), results in a state-of-the-art 32B parameter model on \sbv with 51.9\% pass@1 averaged over three seeds, achieving comparable results to concurrent work (SWE Mirror) while using a trajectory mixture that is 40\% smaller (7k vs. 12k). 
The 2.1\% performance improvement over \basemix cannot be achieved by continued training on other data of the same size  (\basemix+ \done, \basemix+ \dtwo). 
The results suggest that we have found a method of synthetic bug generation that more closely matches the distribution of bugs found in human-authored edits. 
This is particularly visible when comparing \dfour with \dthree. The intentional nature of \dthree does not yield visible gains over \basemix. The improvement w.r.t. to the baseline in Pass$^3$ shows that the model gets more consistent in resolution strategies across seeds.

In both the 32B and 14B settings, we find that performing fine-tuning with RL using \dfour bugs can lead to even better performance (32B: 52.4\%, 14B: 45.27\% on \sbv).
Finally we observe that the best way to use our data is by training the base model with a combined dataset of trajectories, yielding \frogboss{} and \frogmini{}, our best performing 32B and 14B models.

\paragraph{Challenging Splits} 
While SWE-Bench Verified has been extensively studied and tested in the community, there are subsets of the benchmark that drive most of the progress for state-of-the-art models. 
Namely, the Frontier and Challenging problems are problems on which \opus achieves 11\% and 31\% respectively, despite achieving 73.60\% on the entire set of SWE-Bench Verified.
The Hard problems are problems rated by an expert human SWE to require more than one hour to solve with an overall 42.2\% solve rate by \opus. 
\footnote{Subsets from \url{https://huggingface.co/datasets/jatinganhotra/SWE-bench_Verified-discriminative}}
The Multi-File problems are any problems in SWE-Bench Verified that cross more than one file and solved 10.0\% of the time by \opus.  
We report our results on the subsets in \Cref{tab:swebench_harder}.
We find that on all subsets of SWE-Bench Verified that the inclusion of \dfour results in improved performance over \basemix in Pass@1 score averaged over three seeds.
Moreover, including \dfour results in a 1\% improved in the challenging and frontier subsets. 
However, on the multi-file and hard subsets, fine-tuning further on \dtwo results in improved performance over \dfour. 
This may be because the relatively small size of hard and multi-file subsets (40 and 45 respectively) may have contributed to a higher variance result. 
Notably, training with RL on \dfour from \basemix shows improvement over \basemix on challenging and hard problems, but no improvement over \basemix on frontier and multi-file problems.
Moreover, training with RL does not improve over training with SFT on \dfour for any subset. 
This could be because we use the GRPO algorithm that requires a problem to be partially solvable to make progress - if problems are too difficult, the advantage will be zero for those problems.
However, training on more data to get \frogboss{} performs better than any other model on the challenging and frontier datasets, and is on par with our RL fine-tuned model on the hard subset.

\begin{table}[t]
    \centering
    \caption{\textbf{Comparison of fine-tuning on different bug datasets.} 
    We report Pass@1 averaged over three seeds, Pass@3, Pass$^3$ (tasks solved in all three runs), 
    and Pass@Short (best of three with the shortest trajectory).}
    \label{tab:main_result_32b}
    \begin{tabular}{l|cccc}
        \toprule
        \textbf{Model} & \textbf{Pass@1} & \textbf{Pass@3} & \textbf{Pass$^3$} & \textbf{Pass@Short} \\
        \midrule
        \multicolumn{5}{c}{Qwen3-32B Models} \\
        \midrule
        Qwen3-32B & 25.00 & 40.00 & 12.60 & 29.40  \\
        \basemix (SFT)  & 49.87 & 63.80 & 37.00 & 50.20 \\
        \midrule
        \basemix + \done (SFT) & 50.73 & 63.60 & 36.60 & 54.60 \\
        \basemix + \dtwo (SFT) & 50.53 & 64.60 & 36.60 & 52.40 \\ 
        \basemix + \dthree (SFT)  & 49.87 & 65.00 & 33.00 & 49.60 \\
        \basemix + \dfour (SFT) & 51.93 & 64.40 & 39.40 & 54.60 \\
        \midrule
        \basemix + \dfour (RL) & 52.40 & 65.60 & 38.20 & 56.80 \\
        \midrule
        \alldata (\frogboss{}, SFT) & \textbf{54.60} & \textbf{67.40} & \textbf{41.20} & \textbf{56.80} \\
        \bottomrule
        \toprule
        \multicolumn{5}{c}{Qwen3-14B Models} \\
        \midrule
        Qwen3-14B & 18.33 & 30.40 & 8.20 & 24.40  \\
        \basemix (SFT)  & 41.13 & 54.80 & 28.40 & 47.60 \\
        \midrule
        \basemix + \dfour (SFT) & 40.40 & 55.40 & 25.80 & 45.00 \\
        \basemix + \dfour (RL)  & 42.20 & 55.00 & 29.20 & 45.80 \\
        \midrule
        \alldata (\frogmini{}, SFT)  & \textbf{45.27} & \textbf{58.20} & \textbf{32.20} & \textbf{49.00} \\
        \bottomrule
    \end{tabular}
\end{table}


\begin{table}[t]
    \centering
    \caption{\textbf{Results on Harder Subsets of SWE-Bench Verified.} We report the Pass@1 averaged over three seeds. The frontier, challenging, hard, and multi-file subsets contain problems where state-of-the-art closed source models struggle.}
    \begin{tabular}{lccccc}
        \toprule
        & Full & Challenging & Frontier & Hard & Multi-file \\
        Size & 500 & 155 & 95 & 45 & 40 \\ 
        \midrule
        Qwen3-32B & 25.33 & 1.29 & 0.35 & 5.19 & 0.83 \\
        \basemix (SFT) & 49.87 & 3.66 & 1.05 & 12.59 & 0.83 \\
        \midrule
        \basemix + \done (SFT) & 50.73 & 5.38 & 1.75 & 9.63 & 2.50 \\
        \basemix + \dtwo (SFT) & 50.57 & 5.59 & 2.11 & 15.56 & \textbf{2.50} \\
        \midrule
        \basemix + \dfour (SFT) & 51.93 & 6.45 & \textbf{2.81} & 14.07 & 1.67 \\
        \basemix + \dfour (RL)  & 52.40 & 5.81 & 0.35 & \textbf{15.56} & 0.83 \\
        \midrule
        \alldata (\frogboss{}, SFT)  & \textbf{54.60} & \textbf{7.96} & 1.05 & 14.81 & 1.67 \\
        \bottomrule
    \end{tabular}
    \label{tab:swebench_harder}
\end{table}

\paragraph{Impact of Teacher Model}
In addition to \claude, we also attempt to collect agent trajectories using \gptiv and \gptv as LLM backbone. 
We report the statistics of these trajectories in Table~\ref{tab:difficulty_stats}.
Performance wise, \gptv outperforms \claude in all four sets of bugs, while \gptiv struggles to resolve even one third of the bugs, especially on the \dthree and \dfour sets. 
In comparison to \claude, we observe that the GPT models tend to generate significantly less assistant content in association with the tool/function calls. 
This is particularly obvious in the trajectories collected on the \done and \dtwo using \gptv as backbone, on average there is less than one assistant content being generated per trajectory.
This is perhaps because the design of the GPT models intentionally prevents the model from generating too much text when calling the tool.

We observe the lack of assistant content (e.g. think tokens combined with the tool call itself) can significantly hurt the performance of a model fine-tuned on such data, even that the teacher model's performance might be better. 
We trained a Qwen3-32B student model on the successful trajectories collected using the GPT models as backbone (similar setting as \basemix). 
The student models trained on \gptv and \gptiv trajectories result in a success rate of 31.40\% and 21.57\% on \sbv, respectively.
This suggests the assistant content (a summary of the teacher's reasoning) is essential in distilling code repairing skills from teacher models into student models, the assistant content may serve as a Chain-of-Thought \citep{wei2022chain} that more effectively conditions the student model to generate the tool/function calls.
Our observation aligns well with recent reasoning curation work \citep{abdin2025phi,zhao2025learning} where they demonstrate the quality of the reasoning content can be crucial in SFT training in domains such as maths and code generation.

\section{Discussion}

Through our extensive experiments, we have shown the utility of our approach for producing difficult bugs that produce efficient training of SWE agents. 
However, one potential drawback of this method is that it may over time become less effective as a distillation technique if the teacher model (e.g. a large closed source model such as \claude) no longer produces bugs while introducing new features. 
To address this, an avenue for future work could be to use the student model (e.g. a model finetuned for Qwen3-32B) itself to generate the bugs. 
This could result in a pipeline whereby a student model produces both its own training problems as well as training data (such as in an RL loop).

While we observe the proposed approach for bug generation results in diverse and difficult bugs, training future SWE-Agents may call for the ability to generate large number of a a specific type of bugs, in order to improve the agent's capabilities in a specific domain.
This could be achieved by extensions to the core idea presented here, whereby the instructions to the bug-generating agent are suitably modified to direct bug-generation towards specific domains or scenarios.
More generally, the approach of using agents to generate training data can be extended beyond "bugs" to crafting a broad range of scenarios that agents may need to be trained for including  test-generation, code-setup and collaboration.

\clearpage
\bibliography{iclr2026_conference}

\begin{thebibliography}{28}
\providecommand{\natexlab}[1]{#1}
\providecommand{\url}[1]{\texttt{#1}}
\expandafter\ifx\csname urlstyle\endcsname\relax
  \providecommand{\doi}[1]{doi: #1}\else
  \providecommand{\doi}{doi: \begingroup \urlstyle{rm}\Url}\fi

\bibitem[Abdin et~al.(2025)Abdin, Agarwal, Awadallah, Balachandran, Behl, Chen, de~Rosa, Gunasekar, Javaheripi, Joshi, et~al.]{abdin2025phi}
Marah Abdin, Sahaj Agarwal, Ahmed Awadallah, Vidhisha Balachandran, Harkirat Behl, Lingjiao Chen, Gustavo de~Rosa, Suriya Gunasekar, Mojan Javaheripi, Neel Joshi, et~al.
\newblock Phi-4-reasoning technical report.
\newblock \emph{arXiv preprint arXiv:2504.21318}, 2025.

\bibitem[Antoniades et~al.(2024)Antoniades, {\"O}rwall, Zhang, Xie, Goyal, and Wang]{antoniades2024swe}
Antonis Antoniades, Albert {\"O}rwall, Kexun Zhang, Yuxi Xie, Anirudh Goyal, and William Wang.
\newblock Swe-search: Enhancing software agents with monte carlo tree search and iterative refinement.
\newblock \emph{arXiv preprint arXiv:2410.20285}, 2024.

\bibitem[Badertdinov et~al.(2025)Badertdinov, Golubev, Nekrashevich, Shevtsov, Karasik, Andriushchenko, Trofimova, Litvintseva, and Yangel]{badertdinov2025swe}
Ibragim Badertdinov, Alexander Golubev, Maksim Nekrashevich, Anton Shevtsov, Simon Karasik, Andrei Andriushchenko, Maria Trofimova, Daria Litvintseva, and Boris Yangel.
\newblock Swe-rebench: An automated pipeline for task collection and decontaminated evaluation of software engineering agents.
\newblock \emph{arXiv preprint arXiv:2505.20411}, 2025.

\bibitem[FAIR CodeGen~Team(2025)]{cwm2025}
Meta FAIR CodeGen~Team.
\newblock Cwm: An open-weights llm for research on code generation with world models, 2025.
\newblock URL \url{https://ai.meta.com/research/publications/cwm/}.

\bibitem[Gandhi et~al.(2025)Gandhi, Tsay, Ganhotra, Kate, and Rizk]{gandhi2025agents}
Shubham Gandhi, Jason Tsay, Jatin Ganhotra, Kiran Kate, and Yara Rizk.
\newblock When agents go astray: Course-correcting swe agents with prms.
\newblock \emph{arXiv preprint arXiv:2509.02360}, 2025.

\bibitem[Guo et~al.(2025)Guo, Yang, Zhang, Song, Zhang, Xu, Zhu, Ma, Wang, Bi, et~al.]{guo2025deepseek}
Daya Guo, Dejian Yang, Haowei Zhang, Junxiao Song, Ruoyu Zhang, Runxin Xu, Qihao Zhu, Shirong Ma, Peiyi Wang, Xiao Bi, et~al.
\newblock Deepseek-r1: Incentivizing reasoning capability in llms via reinforcement learning.
\newblock \emph{arXiv preprint arXiv:2501.12948}, 2025.

\bibitem[Jain et~al.(2025)Jain, Singh, Shetty, Zheng, Sen, and Stoica]{jain2025r2e}
Naman Jain, Jaskirat Singh, Manish Shetty, Liang Zheng, Koushik Sen, and Ion Stoica.
\newblock R2e-gym: Procedural environments and hybrid verifiers for scaling open-weights swe agents.
\newblock \emph{arXiv preprint arXiv:2504.07164}, 2025.

\bibitem[Jimenez et~al.(2023)Jimenez, Yang, Wettig, Yao, Pei, Press, and Narasimhan]{jimenez2023swe-bench}
Carlos~E Jimenez, John Yang, Alexander Wettig, Shunyu Yao, Kexin Pei, Ofir Press, and Karthik Narasimhan.
\newblock Swe-bench: Can language models resolve real-world github issues?
\newblock \emph{arXiv preprint arXiv:2310.06770}, 2023.

\bibitem[Luo(2025)]{Luo_2025}
Michael Luo.
\newblock Deepswe: Training a fully open-sourced, state-of-the-art coding agent by scaling rl, Jul 2025.
\newblock URL \url{https://www.together.ai/blog/deepswe}.

\bibitem[Luo et~al.(2025)Luo, Jain, Singh, Tan, Patel, Wu, Ariyak, Cai, Venkat, Zhu, Athiwaratkun, Roongta, Zhang, Li, Popa, Sen, and Stoica]{deepswe2025}
Michael Luo, Naman Jain, Jaskirat Singh, Sijun Tan, Ameen Patel, Qingyang Wu, Alpay Ariyak, Colin Cai, Tarun Venkat, Shang Zhu, Ben Athiwaratkun, Manan Roongta, Ce~Zhang, Li~Erran Li, Raluca~Ada Popa, Koushik Sen, and Ion Stoica.
\newblock Deepswe: Training a state-of-the-art coding agent from scratch by scaling rl.
\newblock \url{https://pretty-radio-b75.notion.site/DeepSWE-Training-a-Fully-Open-sourced-State-of-the-Art-Coding-Agent-by-Scaling-RL-22281902c1468193aabbe9a8c59bbe33}, 2025.
\newblock Notion Blog.

\bibitem[Ma et~al.(2024)Ma, Cao, Cao, Zhang, Chen, Liu, Liu, Li, Huang, and Li]{ma2024lingma}
Yingwei Ma, Rongyu Cao, Yongchang Cao, Yue Zhang, Jue Chen, Yibo Liu, Yuchen Liu, Binhua Li, Fei Huang, and Yongbin Li.
\newblock Lingma swe-gpt: An open development-process-centric language model for automated software improvement.
\newblock \emph{arXiv preprint arXiv:2411.00622}, 2024.

\bibitem[Pan et~al.(2024)Pan, Wang, Neubig, Jaitly, Ji, Suhr, and Zhang]{pan2024training}
Jiayi Pan, Xingyao Wang, Graham Neubig, Navdeep Jaitly, Heng Ji, Alane Suhr, and Yizhe Zhang.
\newblock Training software engineering agents and verifiers with swe-gym.
\newblock \emph{arXiv preprint arXiv:2412.21139}, 2024.

\bibitem[Tan et~al.(2025)Tan, Luo, Cai, Venkat, Montgomery, Hao, Wu, Balyan, Roongta, Wang, Li, Popa, and Stoica]{rllm2025}
Sijun Tan, Michael Luo, Colin Cai, Tarun Venkat, Kyle Montgomery, Aaron Hao, Tianhao Wu, Arnav Balyan, Manan Roongta, Chenguang Wang, Li~Erran Li, Raluca~Ada Popa, and Ion Stoica.
\newblock rllm: A framework for post-training language agents.
\newblock \url{https://pretty-radio-b75.notion.site/rLLM-A-Framework-for-Post-Training-Language-Agents-21b81902c146819db63cd98a54ba5f31}, 2025.
\newblock Notion Blog.

\bibitem[Team(2025)]{qwen3technicalreport}
Qwen Team.
\newblock Qwen3 technical report, 2025.
\newblock URL \url{https://arxiv.org/abs/2505.09388}.

\bibitem[Wang et~al.(2025)Wang, Zan, Xin, Liu, Wu, and Shen]{wang2025swe}
Junhao Wang, Daoguang Zan, Shulin Xin, Siyao Liu, Yurong Wu, and Kai Shen.
\newblock Swe-mirror: Scaling issue-resolving datasets by mirroring issues across repositories.
\newblock \emph{arXiv preprint arXiv:2509.08724}, 2025.

\bibitem[Wang et~al.(2024)Wang, Li, Song, Xu, Tang, Zhuge, Pan, Song, Li, Singh, et~al.]{wang2024openhands}
Xingyao Wang, Boxuan Li, Yufan Song, Frank~F Xu, Xiangru Tang, Mingchen Zhuge, Jiayi Pan, Yueqi Song, Bowen Li, Jaskirat Singh, et~al.
\newblock Openhands: An open platform for ai software developers as generalist agents.
\newblock \emph{arXiv preprint arXiv:2407.16741}, 2024.

\bibitem[Wei et~al.(2022)Wei, Wang, Schuurmans, Bosma, Xia, Chi, Le, Zhou, et~al.]{wei2022chain}
Jason Wei, Xuezhi Wang, Dale Schuurmans, Maarten Bosma, Fei Xia, Ed~Chi, Quoc~V Le, Denny Zhou, et~al.
\newblock Chain-of-thought prompting elicits reasoning in large language models.
\newblock \emph{Advances in neural information processing systems}, 35:\penalty0 24824--24837, 2022.

\bibitem[Wei et~al.(2025)Wei, Duchenne, Copet, Carbonneaux, Zhang, Fried, Synnaeve, Singh, and Wang]{wei2025swe}
Yuxiang Wei, Olivier Duchenne, Jade Copet, Quentin Carbonneaux, Lingming Zhang, Daniel Fried, Gabriel Synnaeve, Rishabh Singh, and Sida~I Wang.
\newblock Swe-rl: Advancing llm reasoning via reinforcement learning on open software evolution.
\newblock \emph{arXiv preprint arXiv:2502.18449}, 2025.

\bibitem[Xia et~al.(2024)Xia, Deng, Dunn, and Zhang]{xia2024agentless}
Chunqiu~Steven Xia, Yinlin Deng, Soren Dunn, and Lingming Zhang.
\newblock Agentless: Demystifying llm-based software engineering agents.
\newblock \emph{arXiv preprint arXiv:2407.01489}, 2024.

\bibitem[Xie et~al.(2025)Xie, Li, Gao, Du, Lam, Zou, and Chen]{xie2025swe}
Chengxing Xie, Bowen Li, Chang Gao, He~Du, Wai Lam, Difan Zou, and Kai Chen.
\newblock Swe-fixer: Training open-source llms for effective and efficient github issue resolution.
\newblock \emph{arXiv preprint arXiv:2501.05040}, 2025.

\bibitem[Yang et~al.(2024)Yang, Jimenez, Wettig, Lieret, Yao, Narasimhan, and Press]{NEURIPS2024_5a7c9475}
John Yang, Carlos~E. Jimenez, Alexander Wettig, Kilian Lieret, Shunyu Yao, Karthik Narasimhan, and Ofir Press.
\newblock Swe-agent: Agent-computer interfaces enable automated software engineering.
\newblock In A.~Globerson, L.~Mackey, D.~Belgrave, A.~Fan, U.~Paquet, J.~Tomczak, and C.~Zhang (eds.), \emph{Advances in Neural Information Processing Systems}, volume~37, pp.\  50528--50652. Curran Associates, Inc., 2024.
\newblock URL \url{https://proceedings.neurips.cc/paper_files/paper/2024/file/5a7c947568c1b1328ccc5230172e1e7c-Paper-Conference.pdf}.

\bibitem[Yang et~al.(2025)Yang, Leret, Jimenez, Wettig, Khandpur, Zhang, Hui, Press, Schmidt, and Yang]{yang2025swe}
John Yang, Kilian Leret, Carlos~E Jimenez, Alexander Wettig, Kabir Khandpur, Yanzhe Zhang, Binyuan Hui, Ofir Press, Ludwig Schmidt, and Diyi Yang.
\newblock Swe-smith: Scaling data for software engineering agents.
\newblock \emph{arXiv preprint arXiv:2504.21798}, 2025.

\bibitem[Yuan et~al.(2025)Yuan, Moss, Feghali, Singh, Moldavskaya, MacPhee, Caccia, Pereira, Kim, Sordoni, et~al.]{yuan2025debug}
Xingdi Yuan, Morgane~M Moss, Charbel~El Feghali, Chinmay Singh, Darya Moldavskaya, Drew MacPhee, Lucas Caccia, Matheus Pereira, Minseon Kim, Alessandro Sordoni, et~al.
\newblock debug-gym: A text-based environment for interactive debugging.
\newblock \emph{arXiv preprint arXiv:2503.21557}, 2025.

\bibitem[Zeng et~al.(2025{\natexlab{a}})Zeng, Lv, Zheng, Hou, Chen, Xie, Wang, Yin, Zeng, Zhang, et~al.]{zeng2025glm}
Aohan Zeng, Xin Lv, Qinkai Zheng, Zhenyu Hou, Bin Chen, Chengxing Xie, Cunxiang Wang, Da~Yin, Hao Zeng, Jiajie Zhang, et~al.
\newblock Glm-4.5: Agentic, reasoning, and coding (arc) foundation models.
\newblock \emph{arXiv preprint arXiv:2508.06471}, 2025{\natexlab{a}}.

\bibitem[Zeng et~al.(2025{\natexlab{b}})Zeng, Li, Xiao, Li, Liu, Yan, Wei, He, Song, Liu, et~al.]{zeng2025skywork}
Liang Zeng, Yongcong Li, Yuzhen Xiao, Changshi Li, Chris~Yuhao Liu, Rui Yan, Tianwen Wei, Jujie He, Xuchen Song, Yang Liu, et~al.
\newblock Skywork-swe: Unveiling data scaling laws for software engineering in llms.
\newblock \emph{arXiv preprint arXiv:2506.19290}, 2025{\natexlab{b}}.

\bibitem[Zhang et~al.(2025)Zhang, He, Zhang, Kang, Li, Xie, Wang, Wang, Huang, Fu, et~al.]{zhang2025swe}
Linghao Zhang, Shilin He, Chaoyun Zhang, Yu~Kang, Bowen Li, Chengxing Xie, Junhao Wang, Maoquan Wang, Yufan Huang, Shengyu Fu, et~al.
\newblock Swe-bench goes live!
\newblock \emph{arXiv preprint arXiv:2505.23419}, 2025.

\bibitem[Zhao et~al.(2025)Zhao, Caccia, Shi, Kim, Yuan, Xu, C{\^o}t{\'e}, and Sordoni]{zhao2025learning}
Wanru Zhao, Lucas Caccia, Zhengyan Shi, Minseon Kim, Xingdi Yuan, Weijia Xu, Marc-Alexandre C{\^o}t{\'e}, and Alessandro Sordoni.
\newblock Learning to solve complex problems via dataset decomposition.
\newblock In \emph{2nd AI for Math Workshop@ ICML 2025}, 2025.

\bibitem[Zheng et~al.(2024)Zheng, Zhang, Zhang, Ye, Luo, Feng, and Ma]{zheng2024llamafactory}
Yaowei Zheng, Richong Zhang, Junhao Zhang, Yanhan Ye, Zheyan Luo, Zhangchi Feng, and Yongqiang Ma.
\newblock Llamafactory: Unified efficient fine-tuning of 100+ language models.
\newblock \emph{arXiv preprint arXiv:2403.13372}, 2024.

\end{thebibliography}
\bibliographystyle{iclr2026_conference}

\clearpage
\appendix

\section{Agentic Synthetic Bug Generation} \label{sec:buggen_prompts}

\begin{tcblisting}{
  enhanced,
  breakable,
  title={Purposeful Bug Introduction},
  listing only,
  listing options={
    basicstyle=\footnotesize\ttfamily,
    breaklines=true
  }
}
<uploaded_files>
{{working_dir}}
</uploaded_files>
I've uploaded a python code repository in the directory {{working_dir}}.
Your job is to to introduce subtle runtime bugs that cannot be reliably detected through code reading alone and require debugging tools to diagnose.  
The bug you introduce must cause an existing test to fail but should require runtime debugging tools (like pdb, breakpoints, or state inspection) to diagnose. 
It should NOT be detectable through careful code reading or looking at the stacktrace of the failing test alone. 
Focus on runtime state issues, reference sharing, timing dependencies, or complex execution flows that only become apparent during execution.

To this end, some kinds of bugs you might introduce include:
- Create data flow bugs through deep object mutation: Modify nested data structures (like dictionaries within lists within objects) where the mutation path is long and the effect appears far from the cause.
- Implement context-dependent behavior with global state pollution: Use global variables or class-level state that gets modified as a side effect, causing functions to behave differently depending on previous execution history.
- Implement recursive functions with shared mutable state: Use mutable default arguments or class-level variables in recursive functions that accumulate state across different call trees, causing interference between separate recursive operations.
- Create shared reference issues with mutable objects: Use the same mutable object reference across multiple operations without proper copying, causing modifications in one context to unexpectedly affect another (e.g., sharing lists or dictionaries between instances). 
- Introduce accidental state mutations in nested calls: Modify object state unexpectedly deep within a chain of method calls, where the mutation appears unrelated to the method's stated purpose (e.g., a validation method that accidentally modifies the object being validated.

Tips for introducing the bug:
- It should not cause compilation errors.
- It should not be a syntax error.
- It should not modify the documentation significantly.
- It should cause a pre-exisiting test to fail. But the bug should not be easy to diagnose just by looking at the stacktrace of the failing test.
- The root cause should be separated from the symptom manifestation - where the bug occurs should be different from where the error appears.
- The bug maybe a result of edits to multiple function/files which interact in complex ways.
- The bug should require runtime inspection such as stepping through execution with a debugger to trace the actual cause - it cannot be reliably detected through static code analysis alone.
- For functions with complex state or multiple objects, introduce bugs that span multiple method calls or object interactions.
- Focus on bugs that involve shared state, reference aliasing, or side effects that are not immediately obvious but is only visible during execution.
- The bug should require tools like pdb, debugger breakpoints, or runtime state inspection to diagnose effectively.
- Please DO NOT INCLUDE COMMENTS IN THE CODE indicating the bug location or the bug itself.

Follow these steps to introduce the bug:
1. As a first step, it might be a good idea to go over the general structure of the repository.
2. Decide where and what kind of bug you want to introduce.
3. Plan out how you might need to make changes to introduce this bug.
4. Make the changes by editing the relevant parts of the codebase.
5. Make sure that after editing the code to introduce the bug, at least one pre-existing test fails.
6. Make sure that the bug you have introduced cannot be deteced by looking at the code or the stacktrace alone, and it need the use of debugging tools to diagnose.
7. Do not include any comments in the code or point out the bug in any way.
Your thinking should be thorough and so it's fine if it's very long.

\end{tcblisting}

\begin{tcblisting}{
  enhanced,
  breakable,
  title={Feature Addition},
  listing only,
  listing options={
    basicstyle=\footnotesize\ttfamily,
    breaklines=true
  }
}
<uploaded_files>
{{working_dir}}
</uploaded_files>
I've uploaded a python code repository in the directory {{working_dir}}.

Your task is to implement a new feature in this codebase.
First go through the codebase and identify a suitable new feature to add.
Come up with a plan to implement it and then make the necessary changes to the codebase.
You can use the tools provided to edit files, run tests, and submit your changes.
The feature you introduce should not break any existing functionality.
Make sure the edit you make is complex - you should introduce at least two related changes in the codebase in different files.
\end{tcblisting}

\section{Feature Add Example Bugs} \label{sec:featadd_examples}

\begin{tcblisting}{
    enhanced, 
    breakable, 
    title={Type A - API/signature mismatch or backward-compatibility break}, 
    listing only, 
    listing options={
        basicstyle=\footnotesize\ttfamily, 
        breaklines=true
    }
}
ind_available_providers() returns extra provider that breaks expected module list

### Describe the bug

The `find_available_providers()` function is returning an unexpected provider module `faker.providers.technology` that is not part of the expected provider list. This causes issues when comparing the actual providers against the expected set.

### How to Reproduce

Run the following code to see the issue:

```python
from faker.utils import find_available_providers
from importlib import import_module
from faker import META_PROVIDERS_MODULES

modules = [import_module(path) for path in META_PROVIDERS_MODULES]
providers = find_available_providers(modules)
expected_providers = ['faker.providers.address', 'faker.providers.automotive', 'faker.providers.bank', 'faker.providers.barcode', 'faker.providers.color', 'faker.providers.company', 'faker.providers.credit_card', 'faker.providers.currency', 'faker.providers.date_time', 'faker.providers.emoji', 'faker.providers.file', 'faker.providers.geo', 'faker.providers.internet', 'faker.providers.isbn', 'faker.providers.job', 'faker.providers.lorem', 'faker.providers.misc', 'faker.providers.passport', 'faker.providers.person', 'faker.providers.phone_number', 'faker.providers.profile', 'faker.providers.python', 'faker.providers.sbn', 'faker.providers.ssn', 'faker.providers.user_agent']

print("Found providers:", providers)
print("Expected providers:", expected_providers)
print("Match:", providers == expected_providers)
```

**Expected output:**
```
Found providers: ['faker.providers.address', 'faker.providers.automotive', 'faker.providers.bank', 'faker.providers.barcode', 'faker.providers.color', 'faker.providers.company', 'faker.providers.credit_card', 'faker.providers.currency', 'faker.providers.date_time', 'faker.providers.emoji', 'faker.providers.file', 'faker.providers.geo', 'faker.providers.internet', 'faker.providers.isbn', 'faker.providers.job', 'faker.providers.lorem', 'faker.providers.misc', 'faker.providers.passport', 'faker.providers.person', 'faker.providers.phone_number', 'faker.providers.profile', 'faker.providers.python', 'faker.providers.sbn', 'faker.providers.ssn', 'faker.providers.user_agent']
Expected providers: ['faker.providers.address', 'faker.providers.automotive', 'faker.providers.bank', 'faker.providers.barcode', 'faker.providers.color', 'faker.providers.company', 'faker.providers.credit_card', 'faker.providers.currency', 'faker.providers.date_time', 'faker.providers.emoji', 'faker.providers.file', 'faker.providers.geo', 'faker.providers.internet', 'faker.providers.isbn', 'faker.providers.job', 'faker.providers.lorem', 'faker.providers.misc', 'faker.providers.passport', 'faker.providers.person', 'faker.providers.phone_number', 'faker.providers.profile', 'faker.providers.python', 'faker.providers.sbn', 'faker.providers.ssn', 'faker.providers.user_agent']
Match: True
```

**Actual output:**
```
Found providers: ['faker.providers.address', 'faker.providers.automotive', 'faker.providers.bank', 'faker.providers.barcode', 'faker.providers.color', 'faker.providers.company', 'faker.providers.credit_card', 'faker.providers.currency', 'faker.providers.date_time', 'faker.providers.emoji', 'faker.providers.file', 'faker.providers.geo', 'faker.providers.internet', 'faker.providers.isbn', 'faker.providers.job', 'faker.providers.lorem', 'faker.providers.misc', 'faker.providers.passport', 'faker.providers.person', 'faker.providers.phone_number', 'faker.providers.profile', 'faker.providers.python', 'faker.providers.sbn', 'faker.providers.ssn', 'faker.providers.technology', 'faker.providers.user_agent']
Expected providers: ['faker.providers.address', 'faker.providers.automotive', 'faker.providers.bank', 'faker.providers.barcode', 'faker.providers.color', 'faker.providers.company', 'faker.providers.credit_card', 'faker.providers.currency', 'faker.providers.date_time', 'faker.providers.emoji', 'faker.providers.file', 'faker.providers.geo', 'faker.providers.internet', 'faker.providers.isbn', 'faker.providers.job', 'faker.providers.lorem', 'faker.providers.misc', 'faker.providers.passport', 'faker.providers.person', 'faker.providers.phone_number', 'faker.providers.profile', 'faker.providers.python', 'faker.providers.sbn', 'faker.providers.ssn', 'faker.providers.user_agent']
Match: False
```

### Expected behavior

The `find_available_providers()` function should return only the providers that are expected to be in the baseline provider set, without including any additional providers like `faker.providers.technology` that appear to have been added but aren't part of the original expected list.

### Your project

Faker library

### OS

Linux

### Python version

3.10.18

### Additional context

The extra `faker.providers.technology` provider is appearing in the returned list at index 24, shifting the expected `faker.providers.user_agent` to the end. 

\end{tcblisting}

\begin{tcblisting}{
    enhanced, 
    breakable, 
    title={Type B - Logic/conditional bug}, 
    listing only, 
    listing options={
        basicstyle=\footnotesize\ttfamily, 
        breaklines=true
    }
}
`np.row_stack` fails with mixed array shapes in axis=0 mode

Description

When using `np.row_stack` with arrays that have different shapes along non-concatenation axes, the operation fails unexpectedly. This seems to be a regression as the behavior should match NumPy's standard row_stack functionality.

Reproduction:
```python
import autograd.numpy as np

# This should work but fails
arr1 = np.random.random((2, 3))
arr2 = np.random.random((2, 4)) 
arr3 = np.random.random((1, 4))

result = np.row_stack([arr1, (arr2, arr3)])
```

The expected behavior is that `row_stack` should concatenate arrays along axis 0, similar to `vstack`. When passed a list containing both individual arrays and tuples of arrays, it should handle the concatenation properly.

This appears to affect gradient computation as well when used in differentiable contexts.

\end{tcblisting}

\begin{tcblisting}{
    enhanced, 
    breakable, 
    title={Type C - Input Validation, boundary or sentinel handling error}, 
    listing only, 
    listing options={
        basicstyle=\footnotesize\ttfamily, 
        breaklines=true
    }
}
    Option -a doesn't return expected exit code when invalid arguments are provided

Description

The command line option `-a` (which I assume is for setting additional arguments or attributes) isn't behaving correctly when invalid arguments are passed to it. Instead of returning exit code 2 as expected for invalid options, it's returning exit code 0.

This seems to be a regression in the command line argument handling. When you run the command with `-a arg`, it should fail with exit code 2 to indicate invalid usage, but currently it's succeeding (exit code 0).

How to reproduce:
```bash
# This should fail with exit code 2 but returns 0 instead
python -m pygments -a arg
echo $?  # prints 0 but should print 2
```

Expected behavior: The command should exit with code 2 when `-a` is provided with invalid arguments
Actual behavior: The command exits with code 0

This affects any scripts or CI systems that rely on proper exit codes to detect invalid command line usage.
\end{tcblisting}

\begin{tcblisting}{
    enhanced, 
    breakable, 
    title={Type D - Incorrect Argument Forwarding}, 
    listing only, 
    listing options={
        basicstyle=\footnotesize\ttfamily, 
        breaklines=true
    }
}
Custom validator repr() shows incorrect class name when created with validators.create()

Description

When creating a custom validator using `validators.create()`, the `repr()` method shows an incorrect class name. Instead of showing the actual class name, it displays a formatted version based on the version string.

For example:
```python
Validator = validators.create(meta_schema={'$id': 'something'}, version='my version')
validator = Validator({})
print(repr(validator))
# Shows: MyVersionValidator(schema={}, format_checker=None)
# Expected: <actual class name>Validator(schema={}, format_checker=None)
```

The repr output uses "MyVersionValidator" instead of the proper class name, which makes debugging and introspection more difficult when working with custom validators.
\end{tcblisting}

\begin{tcblisting}{
    enhanced, 
    breakable, 
    title={Type E - Missing import/symbol/attribute error example}, 
    listing only, 
    listing options={
        basicstyle=\footnotesize\ttfamily, 
        breaklines=true
    }
}

    ## Pydantic examples in docstrings failing with import errors

Hey folks, I'm running into some issues with the docstring examples in pydantic. It looks like there are some import problems happening when the examples are being executed.

### Describe the bug

When running docstring examples, some of them are failing during execution. The examples seem to be having trouble with imports or module resolution. This is affecting the documentation validation process.

### How to Reproduce

I created a simple script to reproduce the issue:

```python
import pydantic
from pydantic import BaseModel
from typing import TypeVar, Generic

# Try to run some basic pydantic operations that might be in docstrings
T = TypeVar('T')

class MyModel(BaseModel, Generic[T]):
    value: T

# This should work fine normally
model = MyModel[str](value="test")
print(f"Created model: {model}")
```

When this gets executed in the context of docstring evaluation, it seems to run into problems.

### Expected behavior

All docstring examples should execute successfully without import errors or module resolution issues. The examples are supposed to demonstrate proper pydantic usage and should run cleanly.

### Environment

- Python version: 3.10.18
- Pydantic version: Latest from main branch

### Additional context

This seems to be related to how the docstring examples are being evaluated and potentially how modules are being imported during the evaluation process. The issue appears to affect multiple examples across different parts of the codebase.

The problem might be related to the dynamic import system or how the evaluation environment is set up for running the docstring examples.
\end{tcblisting}

\begin{tcblisting}{
    enhanced, 
    breakable, 
    title={Type F - State consistency/bookkeeping/caching bug}, 
    listing only, 
    listing options={
        basicstyle=\footnotesize\ttfamily, 
        breaklines=true
    }
}
**Describe the bug**
Memory leak when checking Union types - objects created in the same scope as `check_type()` calls are not being properly garbage collected, causing reference leaks.

**To Reproduce**
Create a simple test case with an object that should be garbage collected after going out of scope:

```python
from typeguard import check_type
from typing import Union

class TestObject:
    def __del__(self):
        print("Object deleted")

def test_leak():
    obj = TestObject()
    check_type(b'test', Union[str, bytes])
    # Object should be deleted here when function exits

test_leak()
# Expected: "Object deleted" should be printed
# Actual: Nothing is printed, indicating the object wasn't deleted
```

Also reproducible with Python 3.10+ union syntax:
```python
def test_leak_new_syntax():
    obj = TestObject()
    check_type(b'test', str | bytes)
    # Object should be deleted here

test_leak_new_syntax()
```

**Expected behavior**
Objects should be properly garbage collected when they go out of scope, even when `check_type()` is called in the same scope. No memory references should be retained by the type checking machinery.

**Environment info**
- Python version: 3.10+ (affects both typing.Union and new union syntax)
- typeguard version: latest

**Additional context**
This appears to affect both the legacy `typing.Union` syntax and the newer `str | bytes` union syntax introduced in Python 3.10. The issue suggests that the type checking code may be holding onto references that prevent proper cleanup of local objects.
\end{tcblisting}

\begin{tcblisting}{
    enhanced, 
    breakable, 
    title={Type G - Copy Semantics, mutability aliasing, or in-place mutation of inputs}, 
    listing only, 
    listing options={
        basicstyle=\footnotesize\ttfamily, 
        breaklines=true,
    }
}
### Contrast improvements break test with `fail_if_improved` assertion

I ran into an issue where the contrast test is failing with the message "congrats, you improved a contrast! please run ./scripts/update_contrasts.py". This happens when the contrast values for pygments styles have been improved but the test baseline hasn't been updated.

### How to Reproduce

The issue occurs when running the contrast tests and some style has improved contrast ratios compared to the stored baseline values. The test will fail with an assertion error indicating that contrasts have improved.

### Expected behavior

The test should either automatically update the baseline values when improvements are detected, or there should be a clearer way to handle contrast improvements without requiring manual script execution.

### Additional context

The test uses a `fail_if_improved` parameter that's set to `True` by default, which causes the test to fail when contrast values are better than the stored baseline. This seems counterintuitive - improvements in contrast should typically be welcomed rather than causing test failures.

The error message suggests running `./scripts/update_contrasts.py` but this creates friction in the development workflow when contrast improvements happen naturally through code changes.
\end{tcblisting}

\begin{tcblisting}{
    enhanced, 
    breakable, 
    title={Type H - Protocol/spec conformance bug}, 
    listing only, 
    listing options={
        basicstyle=\footnotesize\ttfamily, 
        breaklines=true
    }
}

## Bug report

The `tldextract` function and `TLDExtract.extract_str`/`TLDExtract.extract_urllib` methods are failing doctest validation. This appears to be related to how the doctests are being processed or executed.

When running the full test suite, three doctest failures occur:

```
FAILED tldextract/tldextract.py::tldextract.tldextract
FAILED tldextract/tldextract.py::tldextract.tldextract.TLDExtract.extract_str  
FAILED tldextract/tldextract.py::tldextract.tldextract.TLDExtract.extract_urllib
```

The doctests in the main `tldextract` function and the `TLDExtract` class methods are not passing validation, while all other regular unit tests continue to pass successfully.

This suggests there may be an issue with the expected output formatting in the docstrings or how the doctest runner is interpreting the examples. The functionality itself seems to work correctly based on the passing unit tests, but the embedded documentation examples are failing validation.
\end{tcblisting}

\begin{tcblisting}{
    enhanced, 
    breakable, 
    title={Type I - Resource Mishandling Issue}, 
    listing only, 
    listing options={
        basicstyle=\footnotesize\ttfamily, 
        breaklines=true
    }
}

When calling the `navigate()` method on a `URL` object with `None` as the path parameter, it doesn't behave as expected in certain scenarios. This seems to affect URL path resolution when dealing with base URLs that have trailing paths.

Here's a minimal reproduction:

```python
from boltons.urlutils import URL

# This works as expected
url = URL('https://host/a/')
result = url.navigate('b')
print(f"Expected: https://host/a/b, Got: {result.to_text()}")

# This doesn't work correctly  
url = URL('https://host/a')
result = url.navigate(None).navigate('b')
print(f"Expected: https://host/b, Got: {result.to_text()}")

url = URL('https://host/a/')
result = url.navigate(None).navigate('b') 
print(f"Expected: https://host/a/b, Got: {result.to_text()}")
```

Expected behavior:
...

The issue appears to be in how `navigate()` handles `None` paths when resolving relative URLs. The method should properly handle the case where `None` is passed as a path parameter and maintain correct URL resolution behavior for subsequent chained `navigate()` calls.

This affects URL manipulation when programmatically building URLs where the path might be conditionally `None`.
\end{tcblisting}

\section{Bug Categorisation} \label{sec:appendix_bugcat}

We use a hierarchical summarisation strategy to come up with bug types to categorise bugs.
Bugs from all datasets are pooled togethers and an LLM is used to come up with summaries of individual bugs along with potential bug types.
These summaries are grouped together and further summarised.
We continue this process and obtain the following ten bug categories -

\begin{tcblisting}{
    enhanced, 
    breakable, 
    title={Bug Category Descriptions}, 
    listing only, 
    listing options={
        basicstyle=\footnotesize\ttfamily, 
        breaklines=true
    }
}
A: API/signature mismatch or backward-compatibility break
  - Description: Public interfaces change or fail to accept/forward expected parameters; options no longer propagated; removed/renamed methods.
  - Signals: TypeError for unexpected/unknown keyword, missing method attribute, inability to customize behavior that used to work.
  - Common fixes: Align signatures across layers, add/propagate parameters, restore deprecated shims or document breaking changes.

B: Logic/conditional bug
  - Description: Incorrect branching, inverted predicates, off-by-one comparisons, or misplaced conditions that alter behavior.
  - Signals: Wrong results for specific ranges/cases; behavior flips when a flag toggles; regression tied to a refactor of if/else logic.
  - Common fixes: Correct predicates/ordering; add minimal repro tests around boundary values and both branches.

C: Input validation, boundary, or sentinel handling error
  - Description: Valid inputs rejected or invalid accepted; special values (NaN/None/NA/masked) mishandled due to comparison/identity semantics.
  - Signals: Edge cases fail while common cases pass; inconsistent behavior with empty inputs or special sentinels.
  - Common fixes: Validate before use; use library-appropriate checks for sentinels; add edge/empty-case tests.

D: Incorrect argument forwarding, constructor, or inheritance contract break
  - Description: Subclasses pass wrong args to super, fail to call base initializer, or expose mismatched signatures.
  - Signals: TypeError/AttributeError during object creation; missing base attributes; framework hooks not invoked.
  - Common fixes: Align constructor signatures; call super() correctly; set attributes after base init; stop forwarding unsupported args.

E: Missing import/symbol/attribute error
  - Description: Required names removed or not imported after refactor; attributes expected by callers no longer present.
  - Signals: NameError/AttributeError at runtime; module-level failures on import.
  - Common fixes: Restore or re-export symbols; update imports; add import-time tests.

F: State consistency/bookkeeping/caching bug
  - Description: Shared or stale state corrupts behavior across calls/instances; counters/heaps not updated; cache keys too coarse.
  - Signals: Nondeterministic results; memory growth; behavior depends on call order; leaked/stale entries.
  - Common fixes: Use per-call/per-instance state; fix increment/decrement paths; design proper cache keys; add isolation/concurrency tests.

G: Copy semantics, mutability aliasing, or in-place mutation of inputs
  - Description: Wrong choice of shallow/deep copy; shared mutable defaults; functions mutate caller-provided objects.
  - Signals: Changes in one consumer affect another; unexpected side effects; duplicated or missing internal state.
  - Common fixes: Avoid mutating inputs; pick correct copy depth; use default_factory for mutables; return defensive copies.

H: Protocol/spec conformance bug
  - Description: Behavior violates external specs (HTTP, OAuth, data interchange) or expected wire formats.
  - Signals: Clients reject responses; strict parsers fail; tests asserting spec rules break (e.g., HTTP HEAD body handling).
  - Common fixes: Implement per spec; adjust emission/validation logic; add conformance tests.

I: IO/filesystem/resource handling bug
  - Description: Incorrect handling of paths/streams/resources; special-case short-circuits skip real writes; missing directory creation.
  - Signals: Truncated output; OSError/FileNotFoundError; behavior differs between stdout vs file.
  - Common fixes: Ensure normal write paths execute; create/check dirs; close/flush properly; test both special and normal streams.

J: Security/sensitive-data leakage due to logic oversight
  - Description: Credentials/headers applied too broadly (e.g., to all domains) or without proper scoping/validation.
  - Signals: Tokens sent to unintended endpoints; security reviews flag over-permissive defaults.
  - Common fixes: Scope credentials to allowed domains; enforce whitelists; secure defaults; add security-focused tests.
\end{tcblisting}

We use the following prompt to categorise individual bugs into one of these buckets -

\begin{tcblisting}{
    enhanced, 
    breakable, 
    title={Bug Categorisation Prompt}, 
    listing only, 
    listing options={
        basicstyle=\footnotesize\ttfamily, 
        breaklines=true
    }
}
Your task is to categorise a provided bug into a set of given bug types.

Here are the guidelines on the bug types - 
{guide}

Here is the bug the needs to be categorised -

<problem_description>
{ps}
</problem_description>

<patch>
{patch}
</patch>

Your response should be in xml format:
<reasoning>
Thinking about which categories that the given bug falls into.
</reasoning>
<category>
Alphabet code of category that bug falls into.
</category>
\end{tcblisting}

\section{Reinforcement Learning} \label{sec:reinforcement_learning}

\begin{figure}[h]
    \centering
    \includegraphics[width=0.6\linewidth]{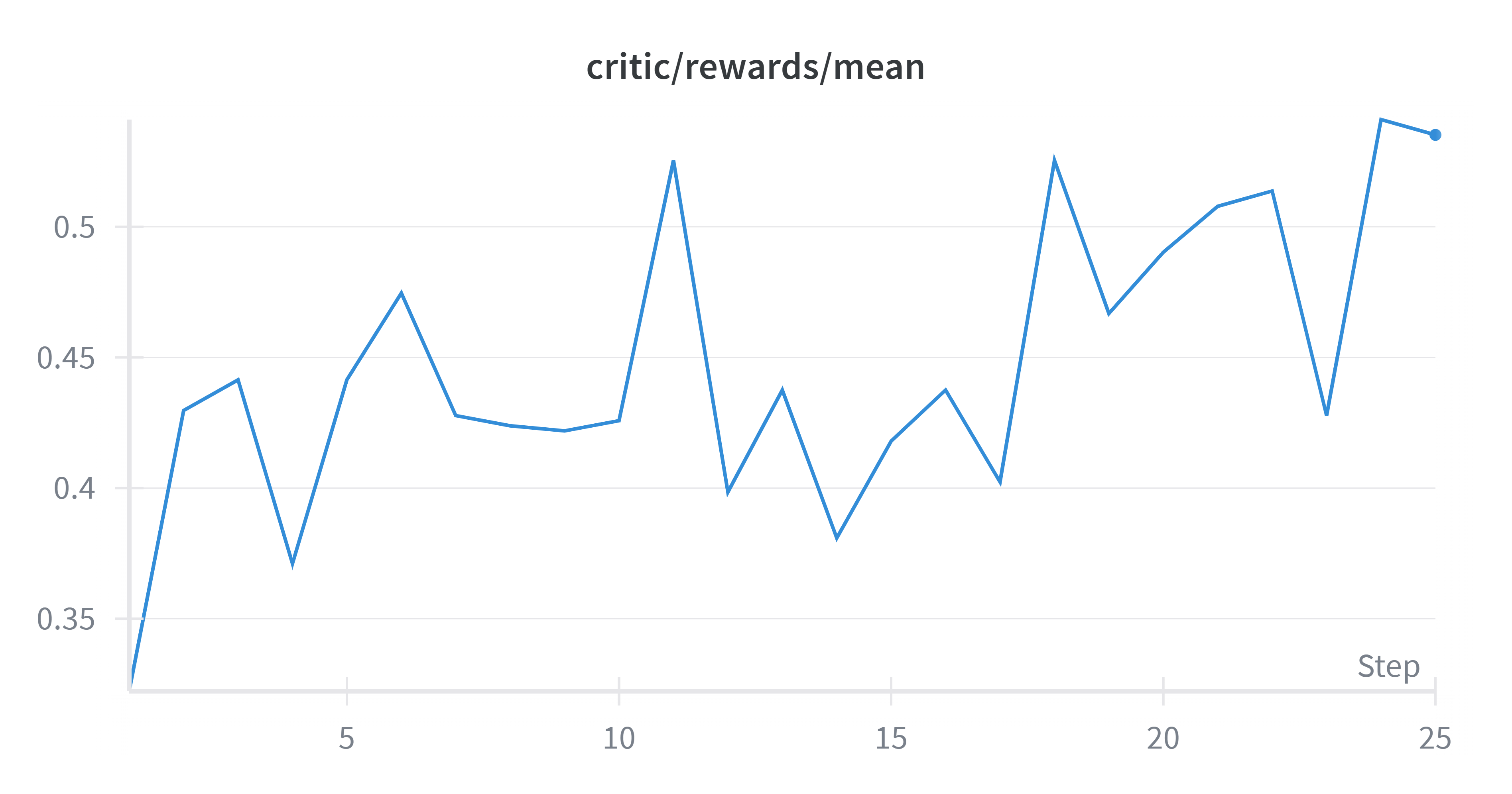}
    \caption{Training reward for reinforcement learning run averaged over batch.
    We train our reinforcement learning from a base model trained on Qwen3-32B on \basemix for 25 steps.
    Each step consists of 64 problems sampled randomly from \dfour and 8 rollouts per problem
    Notably, \claude achieves a 41.4\% total performance on the \dfour dataset, but during our training process during the 25th step, the RL process has over 50\% training reward. 
    This indicates that the RL model may be better at \dfour bugs in general over the course of the training process}
    \label{fig:rl_score}
\end{figure}

To train our model with reinforcement learning, we use the rllm framework \cite{rllm2025}, an open source paradigm to train reinforcement learning. 
Previously, this framework was used to train DeepSWE \cite{deepswe2025} which achieved an overall performance of 41.0\% on \sbv. 
Our training recipe shows an 11.0\% performance improvement over this previous result by bootstrapping from a distilled model with SFT. 
The main changes in hyperparameter between DeepSWE and our model is the use of 100 max steps and 64k context length. 
Because the base SFT checkpoint that we use was trained with 100 steps and used a 64k context length to filter to successful trajectories, we believe that this lack of change in paradigm shift is what led the reinforcement learning to train better in this scenario. 

Note that when we run with the reinforcement learning we use a max context length of 64k and trim to 32k. 
We tried one run where we used a max context length of 32k and max steps of 50 and were able to make progress on the training reward but saw a decrease in performance on the evaluation reward on \sbv. 
Moreover, we tried another run where we finetuned from the base model on a mixture of \done and \dfour bugs but found that the reward was not increasing quickly enough.
Contrary to the advice found in \cite{deepswe2025}, we were able to get a 2.5\% improvement over our base SFT model and achieve state-of-the-art results using an SFT+RL paradigm.

\begin{table}[t]
    \centering
    \caption{Hyperparameters for our Reinforcement Learning Run}
    \begin{tabular}{c|cc|}
         \toprule
         Hyperparameter      & Value \\
         \midrule
         Rollout Temperature & 1.0 \\
         Max Steps           & 100 \\
         Use KL Loss         & False \\
         Train Batch Size    & 64 \\
         Learning Rate       & 1e-6 \\
         PPO Mini Batch Size & 8 \\
         Max Context Length  & 64k \\
         \bottomrule
         
    \end{tabular}
    \label{tab:hyperparameters}
\end{table}

\section{Further Diff Patch Analysis}

In \Cref{tab:lines_of_code}, we analyze how many lines of code, lines of documentation and files were edited and/or created in the diff patch generated by \dtwo, \dthree, and \dfour. 


\begin{table}[t]
    \centering
    \caption{Comparison along axes of how many lines of code were truly created and how many files were edited. \dfour bugs involve many more lines and files edited that \dtwo or \dthree. However, about half of the files edited are new files and half of the lines edited are documentation. }
    \begin{tabular}{c|ccc}
        \toprule
         &  \dtwo & \dthree & \dfour \\
         \midrule
        Avg. Lines of Code &   8.8 & 14.5 & 206.5\\
        Avg. Lines of Documentation  & 4 & 5.8 & 233 \\
        Avg. Files Edited  & 1.18 & 1.50 & 2.19\\
        Avg. Files Created  & 0.004 & 0.09 & 2.43\\
        \bottomrule
        
    \end{tabular}
    \label{tab:lines_of_code}
\end{table}

\end{document}